\documentclass[aps,physrev,twocolumn,superscriptaddress,showkeys]{revtex4-2}

\usepackage{graphicx}
\usepackage{dcolumn}
\usepackage{bm}
\usepackage{comment}
\usepackage{amsmath}
\usepackage{subcaption}
\usepackage{lipsum}
\usepackage{hyperref}
\usepackage{color}
\usepackage{fancyhdr}
\fancypagestyle{plain}{%
  \fancyhf{} 
  \fancyhead[R]{Preprint submitted to Physical Review Fluids}
  
}

\begin{document}

\preprint{Physical Review Fluids}

\thispagestyle{plain}

\title{Timescales and Statistics of Shock-induced Droplet Breakup}

\author{Michael Ullman}
\affiliation{Department of Aerospace Engineering, University of Michigan, Ann Arbor, MI, 48109, USA}

\author{Ral Bielawski}
\affiliation{Department of Mechanical and Aerospace Engineering, University of Central Florida, Orlando, FL, 32816, USA}

\author{Venkat Raman}
\affiliation{Department of Aerospace Engineering, University of Michigan, Ann Arbor, MI, 48109, USA}

\date{\today}


\begin{abstract}
Detonation-based propulsion devices, such as rotating detonation engines (RDEs), must be able to leverage the higher energy densities of liquid fuels in order for them to be utilized in practical contexts. This necessitates a comprehensive understanding of the physical processes and timescales that dictate the shock-induced breakup of liquid droplets. These processes are difficult to probe and quantify experimentally, often limiting measurements to macroscopic properties. Here, fundamental mechanisms in such interactions are elucidated through detailed numerical simulation of Mach 2 and 3 shock waves interacting with 100 $\mu$m water droplets. Using a thermodynamically consistent two-phase formulation with adaptive mesh refinement, the simulations capture droplet surface instabilities and atomization into secondary droplets in great detail. The results show that droplet breakup occurs through a coupled multi-stage process, including droplet flattening, formation of surface instabilities and piercing, and the shedding of secondary droplets from the ligaments of the deformed primary droplet. When considering the dimensionless timescale of Ranger and Nicholls ($\tau$), these processes occur at similar rates for the different shock strengths. The PDFs for the Sauter mean diameters of secondary droplets are bimodal log-normal distributions at $\tau=2$. Modest differences in the degree and rate of liquid mass transfer into droplets less than 5 $\mu$m in diameter are hypothesized to partially derive from differences in droplet surface piercing modes. These results are illustrative of the complex multi-scale processes driving droplet breakup and have implications for the ability of shocks to effectively process liquid fuels.
\end{abstract}


\maketitle


\section{Introduction \label{sec:intro}} 

Over the last decade, significant research efforts have focused on the development of detonation-based propulsion and power generation technologies, including rotating detonation engines (RDEs) \cite{raman_arfmrde, gutmarkpecs}, pulse detonation engines (PDEs) \cite{kailasanath_pdereview, Wolanski_13}, linear or reflective detonation engines \cite{ullman_drone_CF, slabaugh_cf}, and oblique detonation wave engines (ODWEs) \cite{abisleiman2025_odw_cnf, kailasanath2000review, dunlap1958_jpp}. While most studies have focused on gaseous fueling, practical detonation engines must be able to leverage the higher energy densities of liquid fuels. Liquid-fueled detonations require that the fuel atomizes, evaporates, and reacts with the oxidizer prior to reaching the wave-relative sonic plane in order for the resultant heat release to support the wave propagation. Therefore, the timescales associated with droplet breakup and evaporation are of critical importance.

In practical detonation engines, the initial atomization of injected fuel jets is driven by shearing with oxidizer jets. The interaction of the resulting droplet population with the passing detonation wave then facilitates further atomization into smaller droplets and may be responsible for the generation of the vaporized fuel that supports the detonation itself \cite{kateris2023_icders}. Considering that the impact of the detonation wave plays a key role in the droplet breakup, several experimental \cite{nicholls1969_aiaa, duke2023_shockdroplet, salauddin2023_shock_det_droplet, sharma2021_shockdroplet} and computational \cite{bielawski_thesis, bielawski2024analysis, dorschner2020_droplet} studies have considered canonical shock-droplet interactions, wherein one or more initially stationary liquid droplets interact with a traveling shock wave. For shock-droplet interactions relevant to detonations, the Weber numbers are between $10^3 < We < 10^5$, which is within the catastrophic breakup \cite{pilch1987_breakup} or shear-induced entrainment (SIE) regime \cite{sharma2021_shockdroplet, theofanous2008_pof}. Previous works have argued that a combination of Kelvin-Helmholtz (KH) and Rayleigh-Taylor (RT) instabilities drives the breakup processes in this regime \cite{pilch1987_breakup, guildenbecher2009_secondary}, with KH becoming more dominant with increasing $We$ \cite{theofanous2008_pof, theofanous2012_pof}. However, other breakup pathways, such as ballistic-collision-driven evaporation, have also been proposed \cite{kateris2023_icders}.

Because of limitations in spatiotemporal resolution and optical access, shock- and detonation-driven droplet breakup processes are difficult to quantify experimentally and are often limited to macroscopic properties, such as line-of-sight droplet morphology, population, and lifetime \cite{salauddin2023_shock_det_droplet, sharma2021_shockdroplet, burr2024_ilass}. To address this gap, high-fidelity simulations can be leveraged to probe the underlying micro-scale phenomena in greater detail. One common approach for simulating shock- and detonation-droplet interactions \cite{musick2023_eullag, dammati2025_scitech, batista2023_eullag}, as well as full-scale detonation engines \cite{prakash2024_liqRDE, meng2021_eullag, wang2022_eullag}, is to use Euler-Lagrangian formulations with empirical breakup and evaporation models. However, in order to examine breakup processes themselves, numerical approaches that capture or track phasic interfaces are needed. One such interface-capturing approach is the volume of fluid (VOF) method, which has been widely used in simulations of droplet breakup in subsonic and moderate $We$ flows \cite{dorschner2020_droplet, jain2015_secondary, strotos2016_ijmf, poplavski2020_shockdroplet}. On the other hand, comparable simulations in the catastrophic breakup regime are sparse \cite{bielawski_thesis} and often performed in two dimensions to alleviate significant computational cost \cite{bielawski2024analysis, tarey2024_shockdroplet, srinivasan2025_aiaa}. Because of the inherently three-dimensional interface instabilities and shock structures in practical shock-droplet interactions, simulations that account for this dimensionality while capturing both the primary and secondary breakup processes are needed. Interface-capturing schemes, such as VOF, are well suited to this task; however, they do often entail limitations in regard to rates of phase transition. It is common to assume quasi-instantaneous equilibration, where mechanical and thermal equilibrium between the phases are achieved within a given computational cell during a simulation time step \cite{bielawski_thesis, bielawski2024analysis, kuhn2022experimentally}. In reality, phase transitions occur at finite rates due to interactions across sharp phasic interfaces. As such, quasi-instantaneous equilibration routines may lead to artificially accelerated rates of phase transition if the relevant equilibrium should take place over more than one simulation time step. This can be crucial for simulations of liquid-fueled detonations, as the rate of evaporation is a driving parameter for the system. The accuracy and implications of these equilibrium assumptions actively being investigated and may be best addressed by molecular dynamics simulations \cite{kateris2023_icders}.

Noting these points, this work aims to provide further insight into shock-droplet interactions by conducting three-dimensional VOF-based simulations of Mach 2 ($We = 822$) and 3 ($We=3760$) shock waves interacting with water droplets 100 $\mu$m in diameter. Droplet surface instabilities and secondary breakup are captured using adaptive mesh refinement. The results detail the instabilities driving droplet breakup and the subsequent distributions of atomized droplets.


\section{Numerical Methods}

The solver is a multi-phase adaptation of an in-house compressible reacting flow solver \cite{numericsHMM}, which has been extensively used to simulate high-speed flows \cite{ullman2024_cnf_strat, ullman2024_aecs_rde, sharma2024_scitech, sharma2024_proci, abisleiman2025_odw_cnf, rauch2024_scitech}. Block-structured adaptive mesh refinement is handled by the AMReX framework \cite{AMReX_JOSS}. The phases are treated using the volume of fluid (VOF) approach, where $\alpha_l$ and $\alpha_g$ respectively denote the volume fractions of the immiscible liquid and gas phases. The gaseous phase is governed by the ideal gas equation of state, while the liquid phase is governed by the stiffened gas equation of state. As such, the phasic energies are given by:
\begin{equation}
    e_l = \frac{p+\gamma_l \Pi_{l,\infty}}{(\gamma_l -1)\rho_l} + e_{0,l}
    \label{eqn:e_liq_eos}
\end{equation}
\begin{equation}
    e_g = \int_{T_0}^T C_pdT - \frac{p_g}{\rho_g}+\sum_{k=1}^{N_s}\Delta h^0_{f,k}Y_k
    \label{eqn:e_gas_eos}
\end{equation}
where $\gamma_l$, $\Pi_{l,\infty}$, and $e_{0,l}$ are the constant stiffened gas parameters provided by Schmidmayer et al.\ \cite{schmidmayer2020_ecogen} and listed in Table \ref{table:sg_parameters}.

\begin{table}[h]
\centering
\caption{Stiffened gas parameters.}
\begin{tabular}{|c|c|c|c|c|}
\hline
\textbf{$\boldsymbol{c_{v,l}}$} & \textbf{$\boldsymbol{\gamma_l}$} & \textbf{$\boldsymbol{\Pi_{l,\infty}}$} & \textbf{$\boldsymbol{e_{0,l}}$} \\ \hline \hline
1479.48 & 2.82798 & 8.052 $\times 10^8$ & -1.709 $\times 10^7$ \\ \hline
\end{tabular}
\label{table:sg_parameters}
\end{table}

The governing equations were developed by Saurel et al.\ \cite{saurel2009_jcp} and extended by Bielawski et al.\ \cite{bielawski_thesis, bielawski2024analysis} to account for viscous effects and surface tension \cite{schmidmayer2017_jcp}. The mass transport of the mixture and phases is governed by
\begin{equation}
    \frac{\partial \alpha_l}{\partial t} + u_j \frac{\partial \alpha_l}{\partial x_j} = \mathcal{R}_{\alpha_l} (p_l - p_g, T_l-T_g)
    \label{eqn:alpha_l}
\end{equation}
\begin{equation}
    \frac{\partial \alpha_l \rho_l}{\partial t} + \frac{\partial \alpha_l \rho_l u_j}{\partial x_j} = \mathcal{R}_{\rho_l \alpha_l}(p_l-p_g, T_l-T_g)
    \label{eqn:mass_alpha_l}
\end{equation}
\begin{equation}
    \frac{\partial \alpha_g \rho_g}{\partial t} + \frac{\partial \alpha_g \rho_g u_j}{\partial x_j} = \mathcal{R}_{\rho_g \alpha_g}(p_l-p_g, T_l-T_g)
    \label{eqn:mass_alpha_g}
\end{equation}
where the mixture density is given by $\rho = \alpha_l \rho_l + \alpha_g \rho_g$. The conservation equations for momentum and energy are
\begin{equation}
    \frac{\partial \rho u_i}{\partial t} + \frac{\partial \rho u_i u_j}{\partial x_j} + \frac{\partial \alpha_l p_l}{\partial x_i} + \frac{\partial \alpha_g p_g}{\partial x_i} - \frac{\partial \tau_{ij}}{\partial x_j} + \frac{\partial \Omega_{ij}}{\partial x_j} = 0
    \label{eqn:momentum}
\end{equation}

\begin{multline}    
    \frac{\partial (\rho E + \epsilon_\sigma)}{\partial t} + \frac{\partial \rho E u_j}{\partial x_j} + 
    \frac{\partial  \epsilon_\sigma u_j}{\partial x_j} +
    \frac{\partial p u_j}{\partial x_j} - 
    \frac{\partial \tau_{ij} u_i}{\partial x_j} 
    \\
    - \frac{\partial}{\partial x_j} \left( \lambda \frac{\partial T}{\partial x_j} \right) + 
    \frac{\partial u_i \Omega_{ij} }{\partial x_j}
    \\
    = \mathcal{R}_{\rho E} (p_l-p_g, T_l-T_g)
    \label{eqn:energy}
\end{multline}
where
\begin{equation}
    \rho E = \rho_l \alpha_l e_l + \rho_g \alpha_g e_g + \frac{1}{2} \rho u_i u_i
    \label{eqn:rhoE}
\end{equation}
\begin{equation}
    \tau_{ij} = \mu \left( \frac{\partial u_i}{\partial x_j} + \frac{\partial u_j}{\partial x_i} \right) - \frac{2}{3} \mu \frac{\partial u_k}{\partial x_k} \delta_{ij}
    \label{eqn:tau_ij}
\end{equation}
\begin{equation}
    \Omega_{ij} = -\sigma\left(\left(\frac{\partial c}{\partial x_k} \frac{\partial c}{\partial x_k}\right)^{\frac{1}{2}}  \delta_{ij} - \frac{\frac{\partial c}{\partial x_i}\frac{\partial c}{\partial x_j}}{\left(\frac{\partial c}{\partial x_k} \frac{\partial c}{\partial x_k}\right)^{\frac{1}{2}}} \right)
    \label{eqn:omega}
\end{equation}
\begin{equation}
    c = \frac{\alpha_l^{0.1} }{\alpha_l^{0.1} +\left(1 - \alpha_l \right)^{0.1}}
    \label{eqn:color}
\end{equation}

The mixture viscosity $\mu$ and thermal conductivity $\lambda$ are computed using volumetric mixture rules ($\mu = \alpha_l \mu_l + \alpha_g \mu_g$; $\lambda = \alpha_l \lambda_l + \alpha_g \lambda_g$), where the gaseous components are functions of the thermochemical state and the liquid components are held constant at values taken from NIST fluid data at 300 K \cite{nist_fluid_props}. The surface tension coefficient $\sigma$ is also held constant at the reference value for 300 K.

The species conservation within the gas phase is given by
\begin{multline}
    \frac{\partial \rho_g \alpha_g Y_k}{\partial t} + \frac{\partial u_j \rho_g \alpha_g Y_k}{\partial x_j} - \frac{\partial}{\partial x_j} \left( \rho_g D \frac{\partial Y_k}{\partial x_j} \right) - \alpha_g \dot{\omega}_k \\
    = \mathcal{R}_{\rho_g \alpha_g Y_k} (p_l-p_g, T_l-T_g), \quad k=1, \dots, N_s
    \label{eqn:species}
\end{multline}
where $\dot{\omega}_k$ is the chemical source term for the $k$-th species. The relaxation operators $\mathcal{R}$ in Eqs.\ \ref{eqn:alpha_l}-\ref{eqn:species} are each functions of the pressure ($p$) and temperature ($T$) of each of the phases, which are assumed to be in mechanical and thermal equilibrium at phasic interfaces. Within each simulation time step, a relaxation routine is performed such that the relaxation operators tend toward zero, ensuring that both phases have identical pressures and temperatures at interfaces. Unlike in previous works with this solver \cite{bielawski_thesis, sharma2025_aiaa}, this relaxation procedure does not account for phase change. This choice was deliberately made to more easily account for droplet masses during breakup. The reader is referred to Ref.\ \cite{bielawski_thesis} for further details on these relaxation procedures.

The governing equations are solved using a second-order accurate finite-volume discretization, with the Harten-Lax-van Leer-Contact (HLLC) scheme being used for the convective fluxes and central differences being used for the diffusive fluxes. An explicit two-stage Runge-Kutta scheme \cite{shu_rk} is used for the time integration. Interface reconstruction is performed using $\rho$-THINC \cite{garrick2017_rhothinc}. The gas state is handled using the 9 species mechanism of Mueller et al.\ \cite{Mueller_chem}. An in-situ post-processing coloring algorithm, which identifies contiguous regions of liquid, is used to track droplet number, sizes, and locations. Further details on this algorithm and its implementation are provided in Refs.\ \cite{bielawski_thesis, bielawski2024analysis, heinrich2020_coloring}.


\section{Case Configurations}

\begin{figure}[h]
    \begin{center}
    \includegraphics[width=0.9\columnwidth]{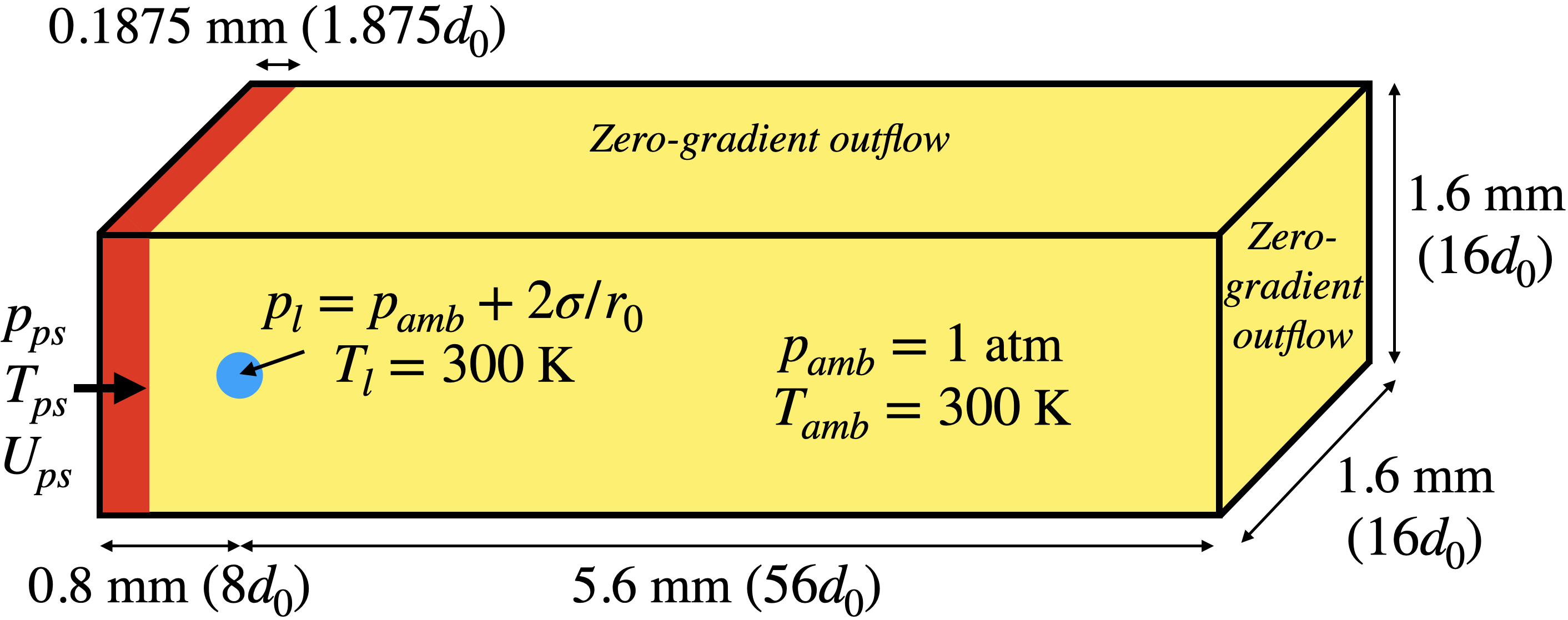}
    \caption{Schematic of initial and boundary conditions.}
    \label{fig:geo_schematic} 
    \end{center}
\end{figure}

A schematic of the case initialization is provided in Fig.\ \ref{fig:geo_schematic}. In both cases, a 100 $\mu$m diameter water droplet is initially suspended in ambient air at 1 atm and 300 K. The droplet is also initialized at 300 K, but with a pressure that includes the added Laplace pressure of $2 \sigma/r_0$. Upstream of the droplet, the inflow boundary uniformly supplies air at the analytic post-shock state ($p_{ps}$, $T_{ps}$, $U_{ps}$) for the desired shock Mach number $M_s$. This creates a roughly uniform flow field behind the shock traveling at the desired speed.

The base computational grid has [256$\times$64$\times$64], yielding a uniform resolution of 25 $\mu$m ($0.25d_0$). Five AMR levels are utilized, each of which divides the grid size along all directions by a factor of 2. This yields a minimum grid resolution of 0.781 $\mu$m ($0.00781d_0$). Adaptive refinement is added using criteria for value and gradient of the liquid volume fraction ($\alpha_l$; $\|\nabla \alpha_l|$), as well as the gradient of the mixture pressure ($|\nabla p|$). These allow the liquid droplets, shocks, and turbulent mixing to be highly resolved. Due to the tagging of regions exceeding the $\alpha_l$ and $\nabla \alpha_l$ thresholds, which grow substantially as the primary droplet atomizes, the maximum cell counts are roughly 980 million for the Mach 2 shock case and 1.15 billion for the Mach 3 shock case. Prior to significant atomization, the cell counts for both cases are between 100-200 million. The cell counts then grow steadily with time as atomization proceeds.

\begin{figure}[!b]
    \begin{center}
    \includegraphics[width=\columnwidth]{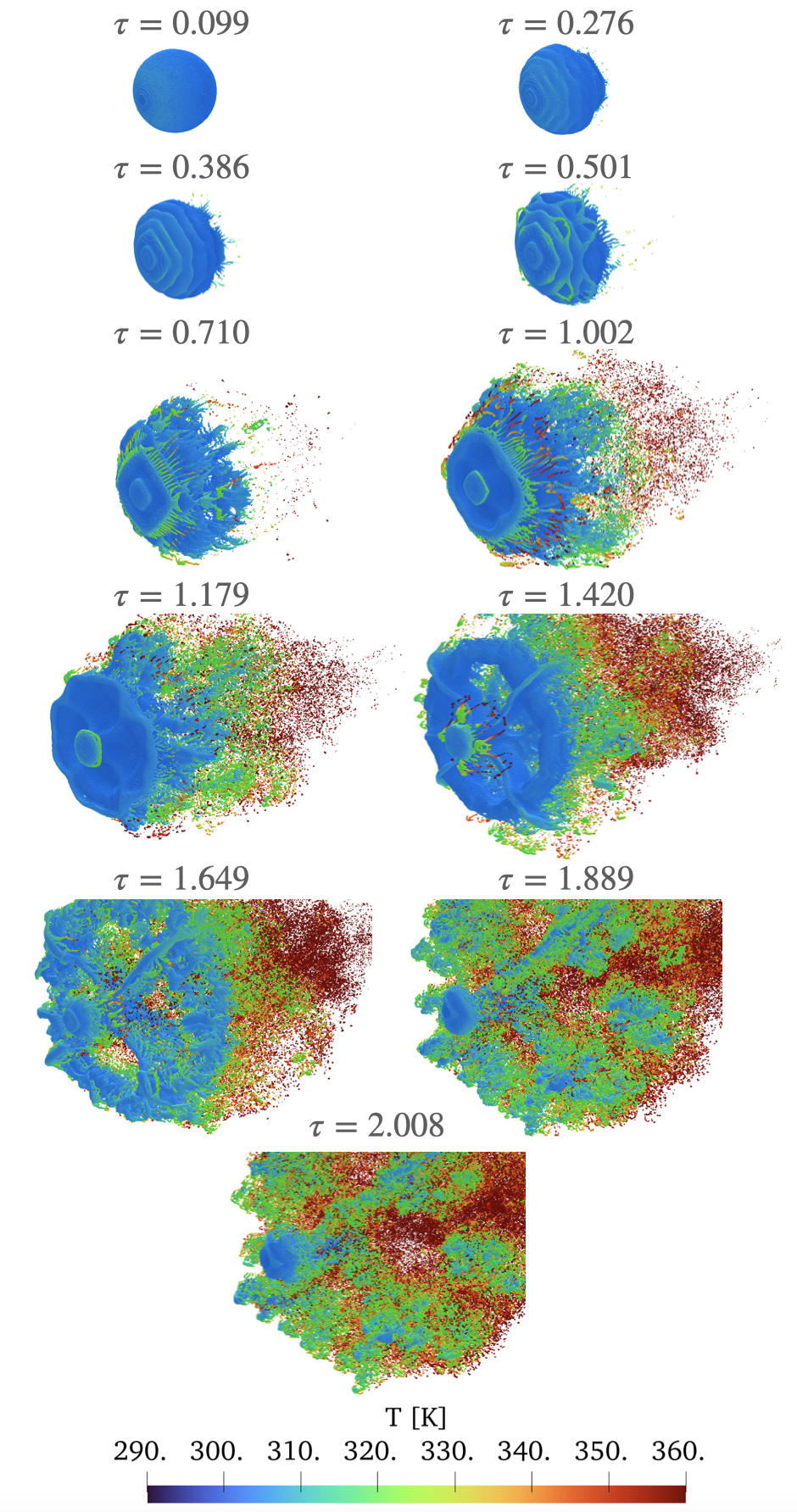}
    \caption{Time sequence of liquid volume fraction contours ($\alpha_l>0.1$) colored by temperature in the $M_s=2$ case.}
    \label{fig:M2_3D_T_sequence} 
    \end{center}
\end{figure}

\begin{figure}[!b]
    \begin{center}
    \includegraphics[width=\columnwidth]{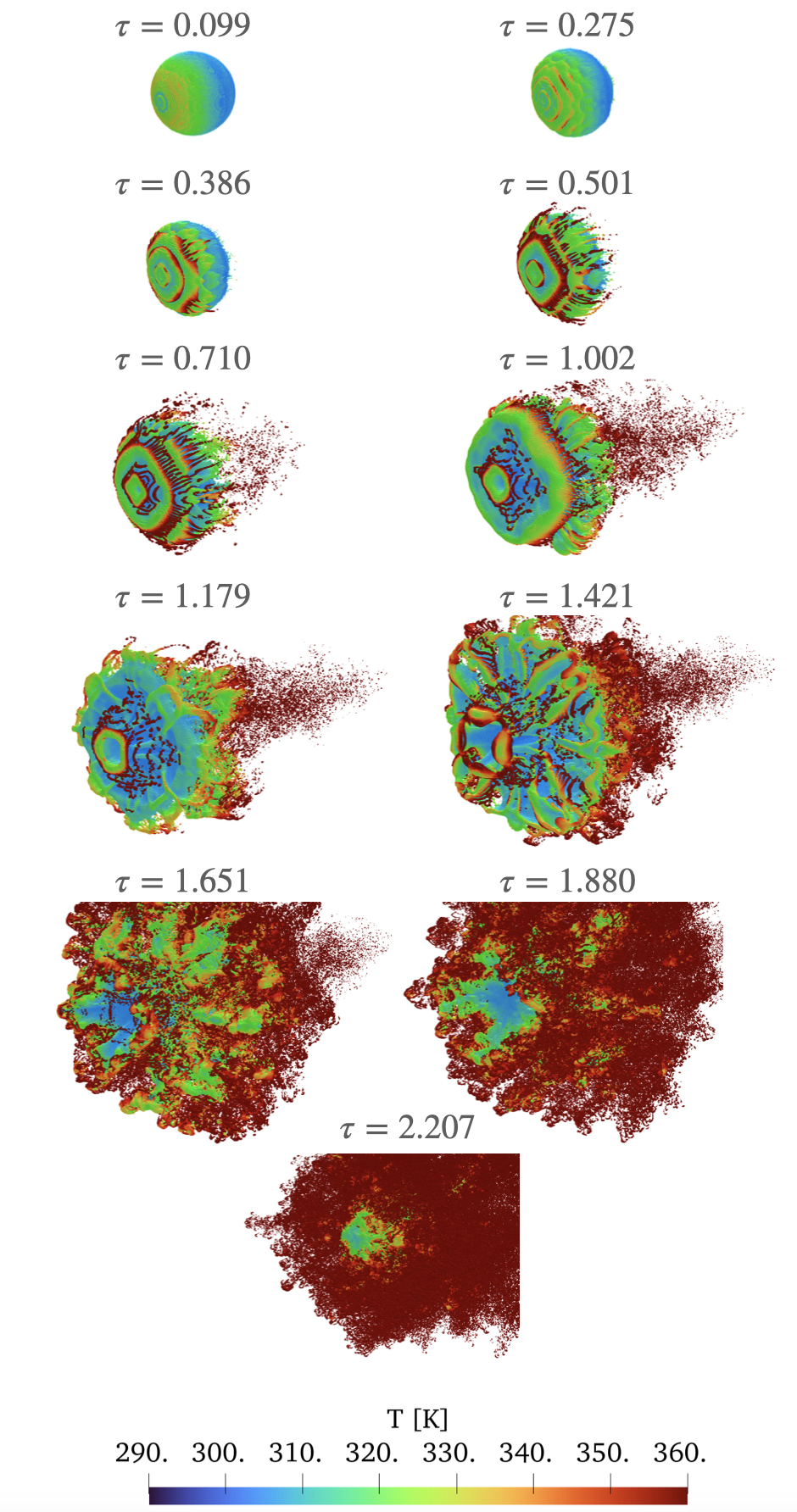}
    \caption{Time sequence of liquid volume fraction contours ($\alpha_l>0.1$) colored by temperature in the $M_s=3$ case.}
    \label{fig:M3_3D_T_sequence} 
    \end{center}
\end{figure}


\section{Results}

\subsection{Instantaneous snapshots}

Figures \ref{fig:M2_3D_T_sequence} and \ref{fig:M3_3D_T_sequence} respectively show time sequences of liquid volume fraction contours colored by temperature in the $M_s=2$ and $M_s=3$ cases. Hereafter, time is expressed using the dimensionless timescale of Ranger and Nicholls \cite{nicholls1969_aiaa}, which is useful for comparing breakup across flow conditions. It is given by
\begin{equation}
    \tau = \frac{t U_r}{d_0} \sqrt{\frac{\rho_g}{\rho_l}}
    \label{eqn:dimensionless_time}
\end{equation}
where $t$ is the physical time and $U_r$ is the relative speed between the droplet and the surrounding gas. The relative speed $U_r$ and gas density $\rho_g$ are here taken as the analytic post-shock conditions for the given shock Mach number $M_s$. When using this dimensionless time, several notable similarities can be observed. The shock has fully traversed the droplet at $\tau \sim 0.01$, but the droplet has not yet begun to deform. Waves of surface instabilities can be seen by $\tau \sim 0.275$, the formation of which coincide with the flattening of the droplet due to the pressure gradient across it. The surface instabilities steadily grow, creating ligaments that start to shed from the primary droplet surface at $\tau \sim 0.38$. These ligaments proliferate and shed small secondary droplets in the wake of the primary droplet. The growth of the surface instabilities eventually facilitates the piercing of the windward side of the droplet at $\tau \sim 1$. This creates even more complex ligament structures as the primary droplet begins to shatter, proceeding until $\tau \sim 2$ when the primary droplet has largely been atomized into a cloud of secondary droplets.

\begin{figure}[!h]
    \begin{center}
    \includegraphics[width=\columnwidth]{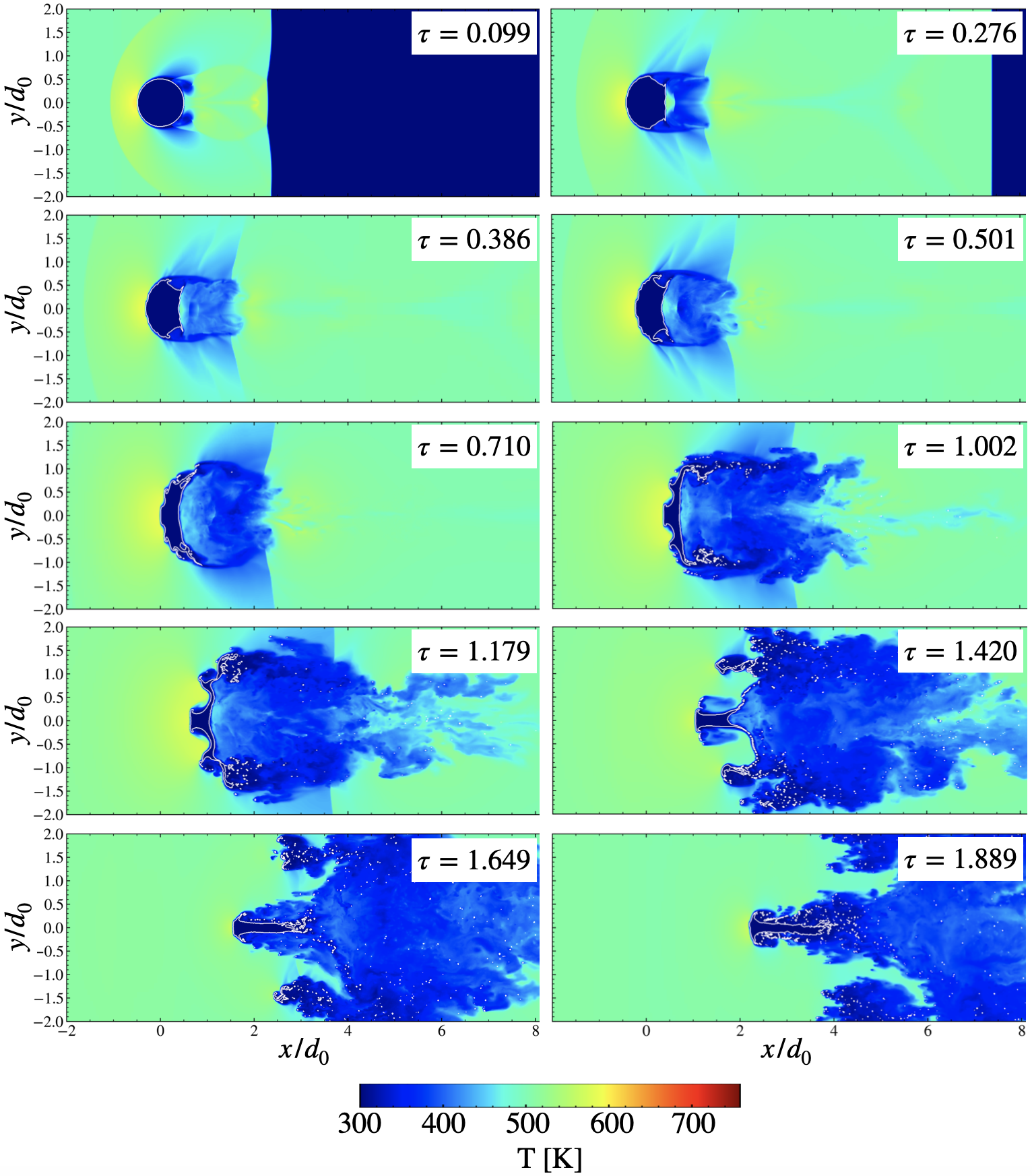}
    \caption{Time sequence of $z$-midplane temperature contours in the $M_s=2$ case. $\alpha_l=1$ isoline marked in white.}
    \label{fig:M2_2D_T_sequence} 
    \end{center}
\end{figure}

\begin{figure}[!h]
    \begin{center}
    \includegraphics[width=\columnwidth]{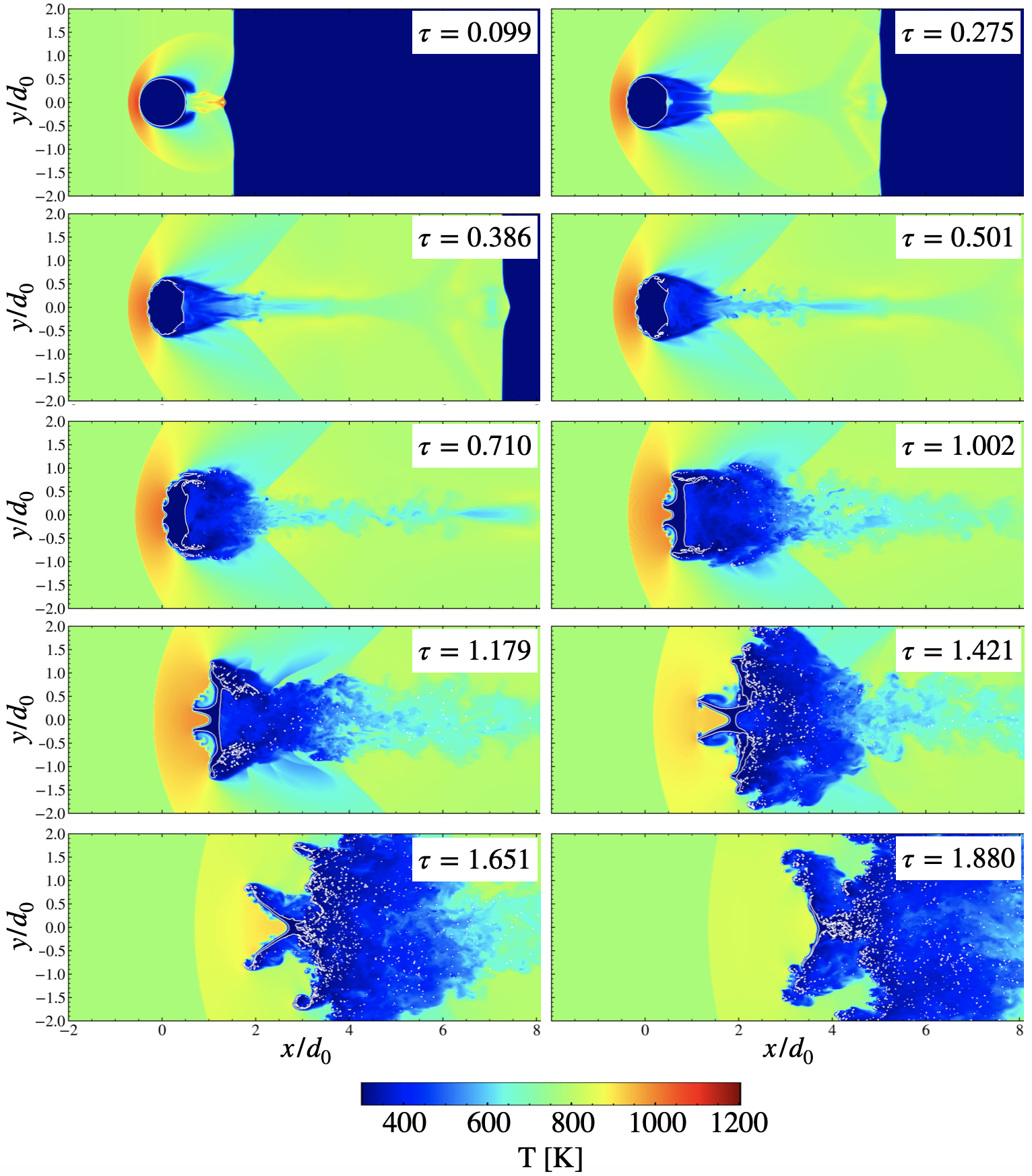}
    \caption{Time sequence of $z$-midplane temperature contours in the $M_s=3$ case. $\alpha_l=1$ isoline marked in white.}
    \label{fig:M3_2D_T_sequence} 
    \end{center}
\end{figure}

The deformation and piercing of the droplets can be seen more clearly in Figs.\ \ref{fig:M2_2D_T_sequence}-\ref{fig:M3_2D_T_sequence}, which show two-dimensional temperature contours along the $z$-midplane for the same time sequences. Here, the droplet morphologies are similar to those previously observed in two-dimensional simulations of shock-droplet interactions \cite{bielawski_thesis, bielawski2024analysis}. As was seen in Figs.\ \ref{fig:M2_3D_T_sequence}-\ref{fig:M3_3D_T_sequence}, secondary ligaments and droplets form along the edges of the primary droplet as it flattens and spreads along its transverse axes. Differences in the droplet morphologies become apparent by $\tau \sim 1$, where the number of surface instabilities on the windward side plays a driving role. The odd-numbered instabilities in the $M_s=2$ case results in a central spike morphology with surface piercing on either side of the spike. On the other hand, the even-numbered instabilities in the $M_s=3$ case results in piercing of the center of the droplet along with the piercing on either side. This leads to the differences in droplet morphology and ligament formation for $\tau > 1$, which may contribute to the differences in secondary droplet formation discussed in the following subsections.


\subsection{Droplet displacement and deformation}

\begin{figure}[!h]
    \centering
    \includegraphics[width=\columnwidth]{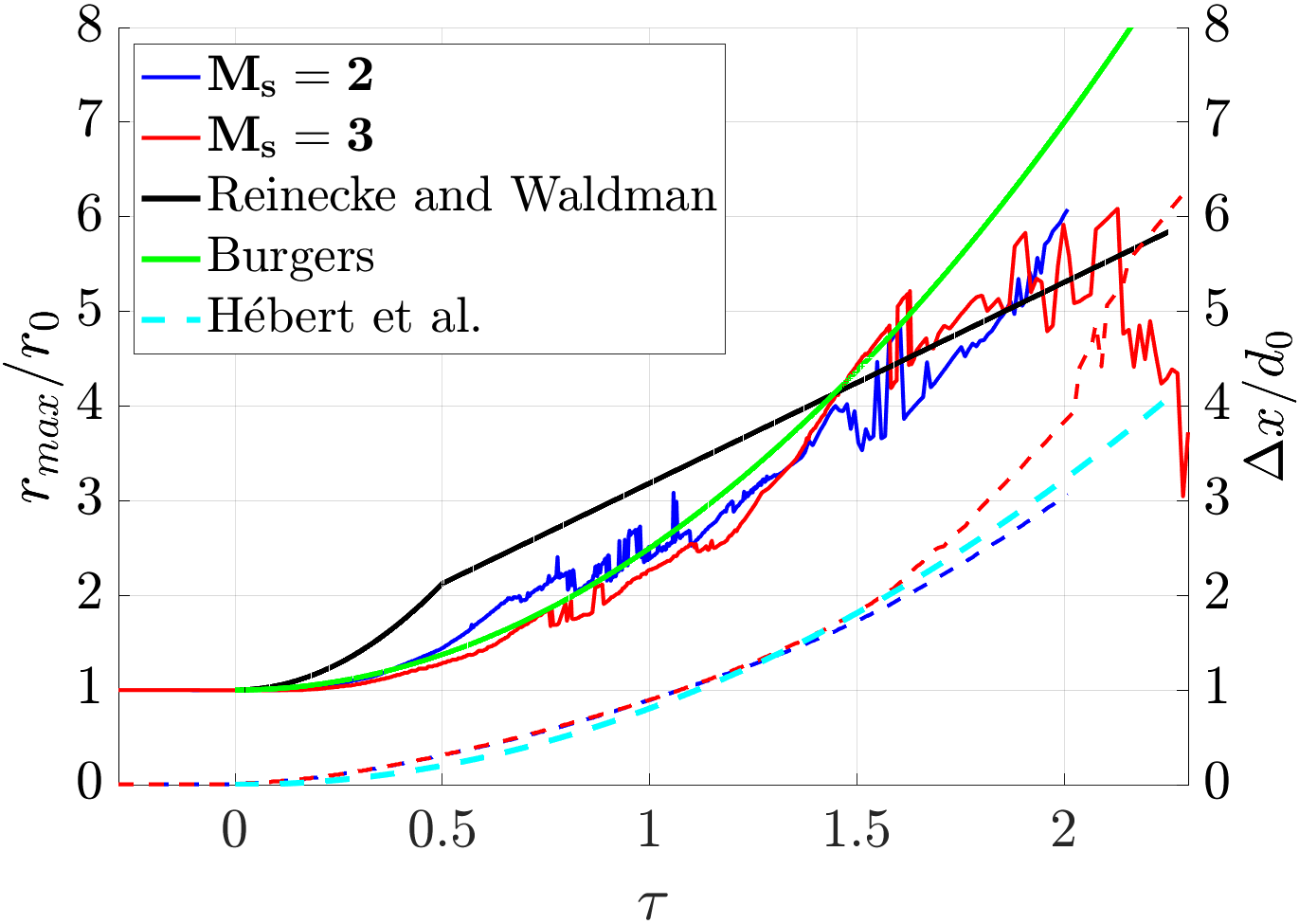}
    \caption{(Left axis; solid lines) Droplet spreading over time. (Right axis; dashed lines) Droplet displacement over time.}
    \label{fig:x_min__r_max__combine__99perc}
\end{figure}

Both the droplet acceleration and deformation have been identified as key parameters influencing the breakup processes \cite{pilch1981acceleration, burr2024_ilass, hebert2020_shockdroplet}. To quantify these, Fig.\ \ref{fig:x_min__r_max__combine__99perc} shows the droplet displacement and spreading as a function nondimensional time. The displacement is computed using the minimum $x$-coordinate at which liquid is identified, and the results are compared to the experimental fit of Hébert et al.\ \cite{hebert2020_shockdroplet}. The spreading is computed as the maximum radial coordinate at which liquid is identified, and the results are compared to the Burgers and Reinecke models presented in Ref.\ \cite{reinecke1975shock}. Following Hébert et al. \cite{hebert2020_shockdroplet}, both quantities are computed by considering computational cells where $\alpha_l \geq 0.99$. This allows $r_{max}$ to be more closely align with the deformation of the contiguous droplet and its ligaments, rather than the spray of atomized secondary droplets.

The dashed profiles in Fig.\ \ref{fig:x_min__r_max__combine__99perc} show that the droplet acceleration is similar in both cases up to $\tau \sim 1.4$, after which point the profiles begin to diverge. This divergence may be expected due to the differences in ligament formation and pierced droplet morphology visible in Figs.\ \ref{fig:M2_2D_T_sequence}-\ref{fig:M3_2D_T_sequence}. The close agreement between these profiles and the experimentally-derived empirical fit of Hébert et al.\ \cite{hebert2020_shockdroplet} is an encouraging validation of the current simulations and provides further confirmation that the initial droplet acceleration is roughly constant with the nondimensional time $\tau$ \cite{mizuno2022deformation, meng2015numerical, bielawski_thesis}.

The solid profiles in Fig.\ \ref{fig:x_min__r_max__combine__99perc} show that similar droplet deformation is observed in both cases up to the conclusion of the $M_s=2$ simulation at $\tau \sim 2$. This self-similarity in droplet deformation for different shock Mach numbers has been noted in several previous experimental works \cite{schroeder2024_pci, nicholls1969_aiaa, reinecke1969experiments, reinecke1969experiments, wierzba1988experimental}. The deformation is initially well-approximated by the Burgers solution for the flattening of a sphere subjected to a uniformly distributed stagnation pressure \cite{engel1958fragmentation}. Once the pierced droplets begin to fragment at $\tau \sim 1.4$, the radial spread of the secondary droplets is better described as an atomized cloud, rather than a deformed sphere. Correspondingly, the radial spread more closely matches the steady-state solution of Reinecke and Walden for a disk-shaped droplet \cite{reinecke1975shock}. These authors found this solution to be in good agreement with several experiments that measured the lateral spread of the atomized secondary droplet cloud \cite{reinecke1975shock, reinecke1970_report}. As such, these results provide further experimental validation for the current simulations.


\subsection{Droplet size distributions}

To quantify the breakup processes, droplet size statistics were computed through time. Following the aforementioned coloring algorithm, a ``droplet" here refers to a contiguous region of liquid. Because of the deformation of these liquid regions, the diameter of a droplet in this context refers to its Sauter mean diameter computed from the volume of the liquid region.

\begin{figure}[!h]
     \centering
     \begin{subfigure}[b]{\columnwidth}
         \centering
         \includegraphics[width=\columnwidth]{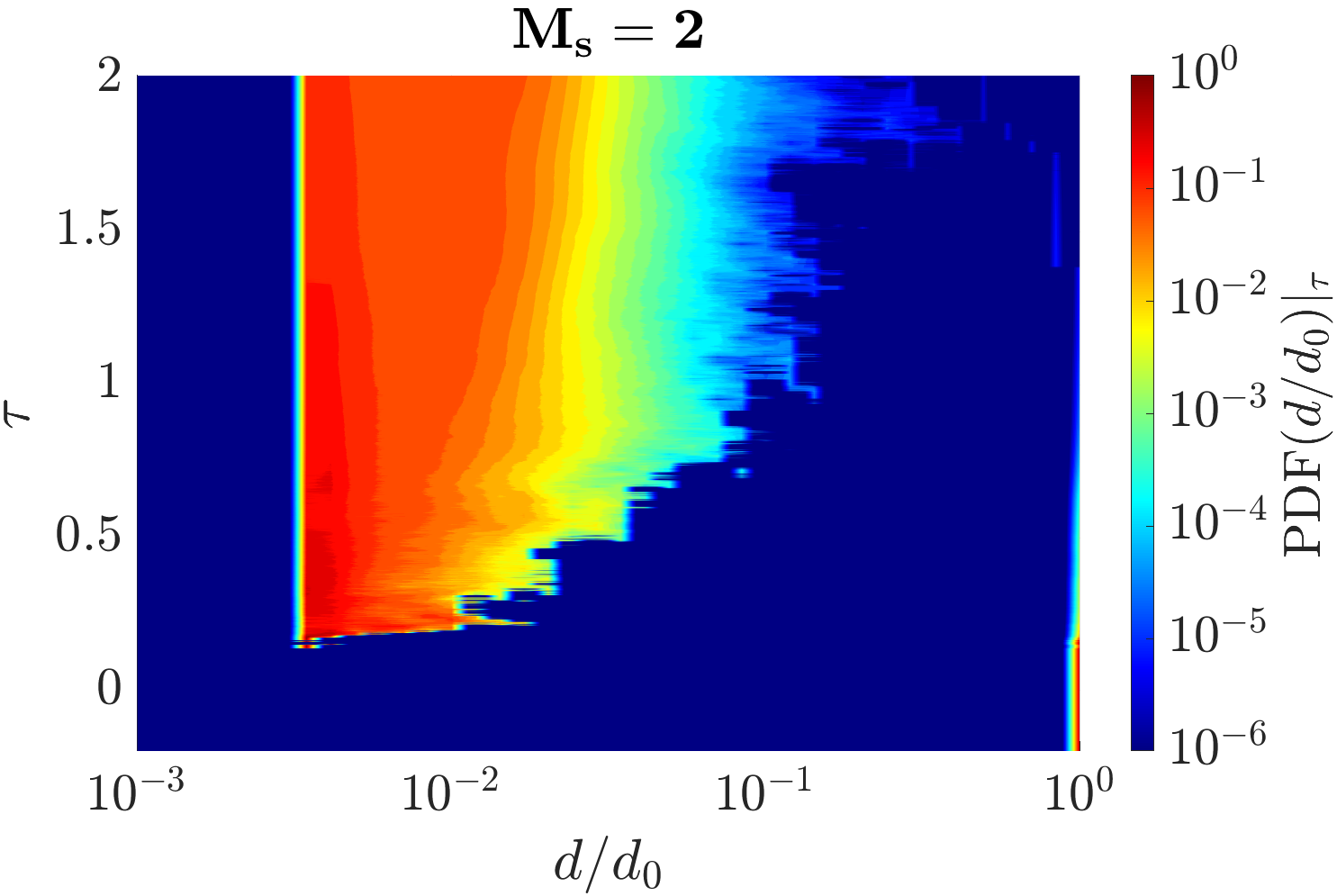} 
     \end{subfigure}
     \\
     \begin{subfigure}[b]{\columnwidth}
         \centering
        \includegraphics[width=\columnwidth]{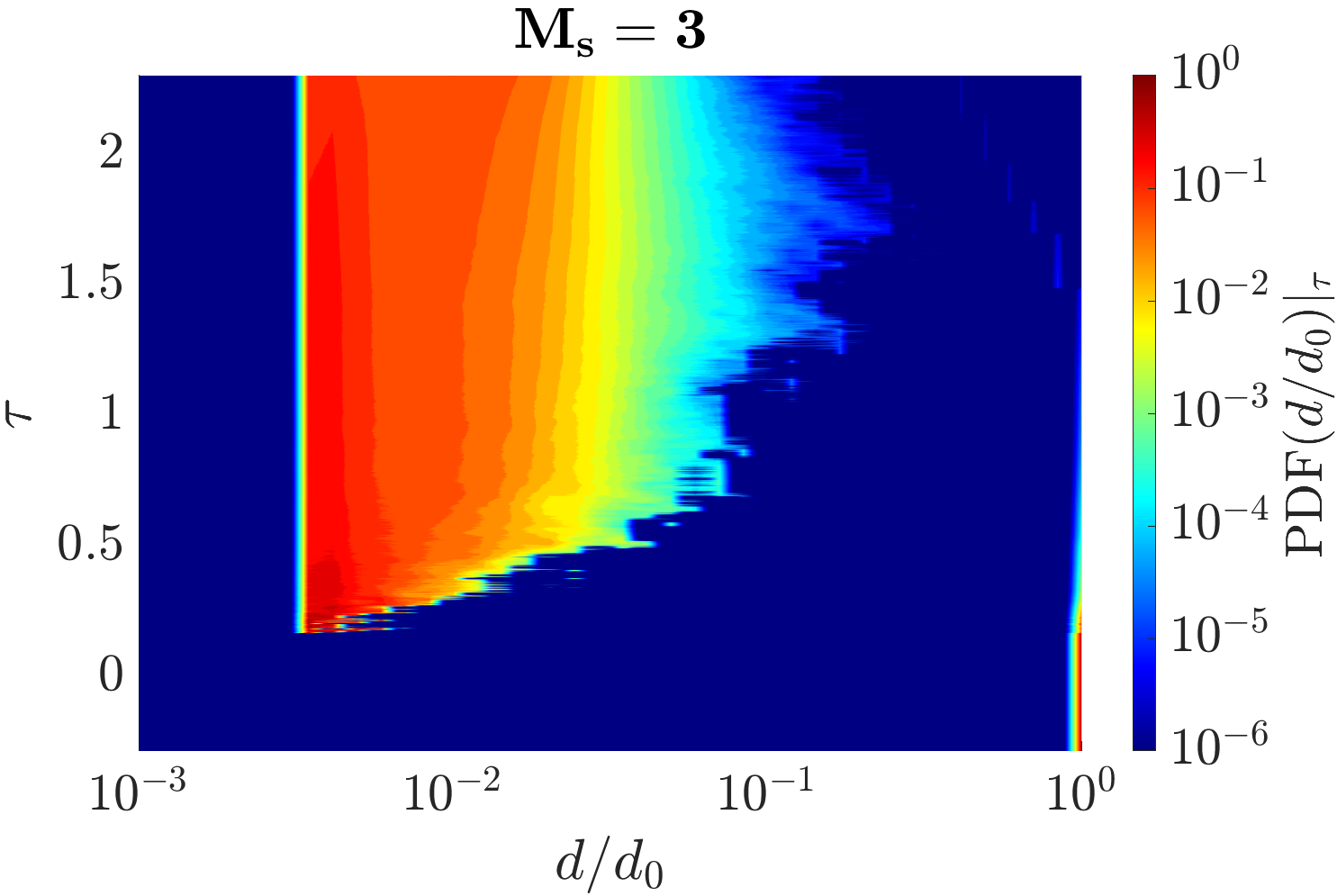} 
     \end{subfigure}
    \caption{PDFs of droplet diameters versus nondimensional time. Normalizations for PDFs are performed for constant $\tau$.}
    \label{fig:d_pdf_surf_const_taus}
\end{figure}

The probability density functions (PDFs) of droplet diameter through time are plotted in Fig.\ \ref{fig:d_pdf_surf_const_taus}. Here, the droplet diameter is normalized by the initial diameter $d_0$, and the PDF is computed independently at each time $\tau$. As such, integrating the PDFs along the $d$ axis yields 1 for each $\tau$. The sharp discontinuities in the PDFs for the smallest $d$ occur due to the smallest grid size, which sets a minimum bound on the resolvable droplet size.

The persistence of the primary droplets can be slivers of the PDFs at $d/d_0=1$ for $\tau<0.15$. After this, secondary droplets with diameters 0.3-1\% of the initial droplet diameter remain the most numerous for the remainder of the simulation durations. However, as the primary droplets are pierced and larger ligaments break off, larger secondary droplets between 1-10\% of $d_0$ become more plentiful. These larger secondary droplets themselves break apart with time, facilitating a cascade of secondary droplet formation and the eventual complete atomization of the primary droplet.

\begin{figure}[!h]
     \centering
     \begin{subfigure}[b]{\columnwidth}
         \centering
         \includegraphics[width=\columnwidth]{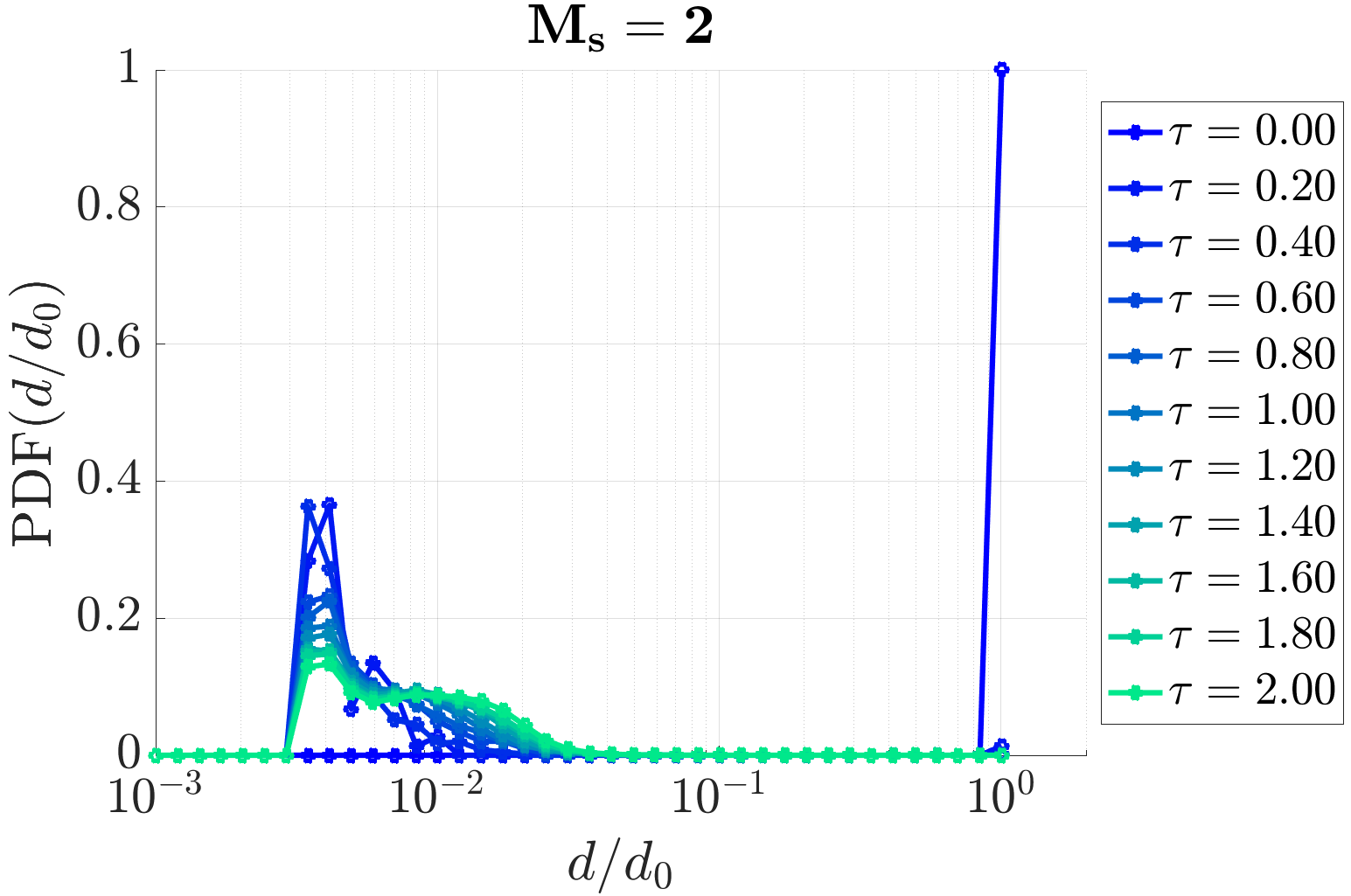} 
     \end{subfigure}
     \\
     \begin{subfigure}[b]{\columnwidth}
         \centering
        \includegraphics[width=\columnwidth]{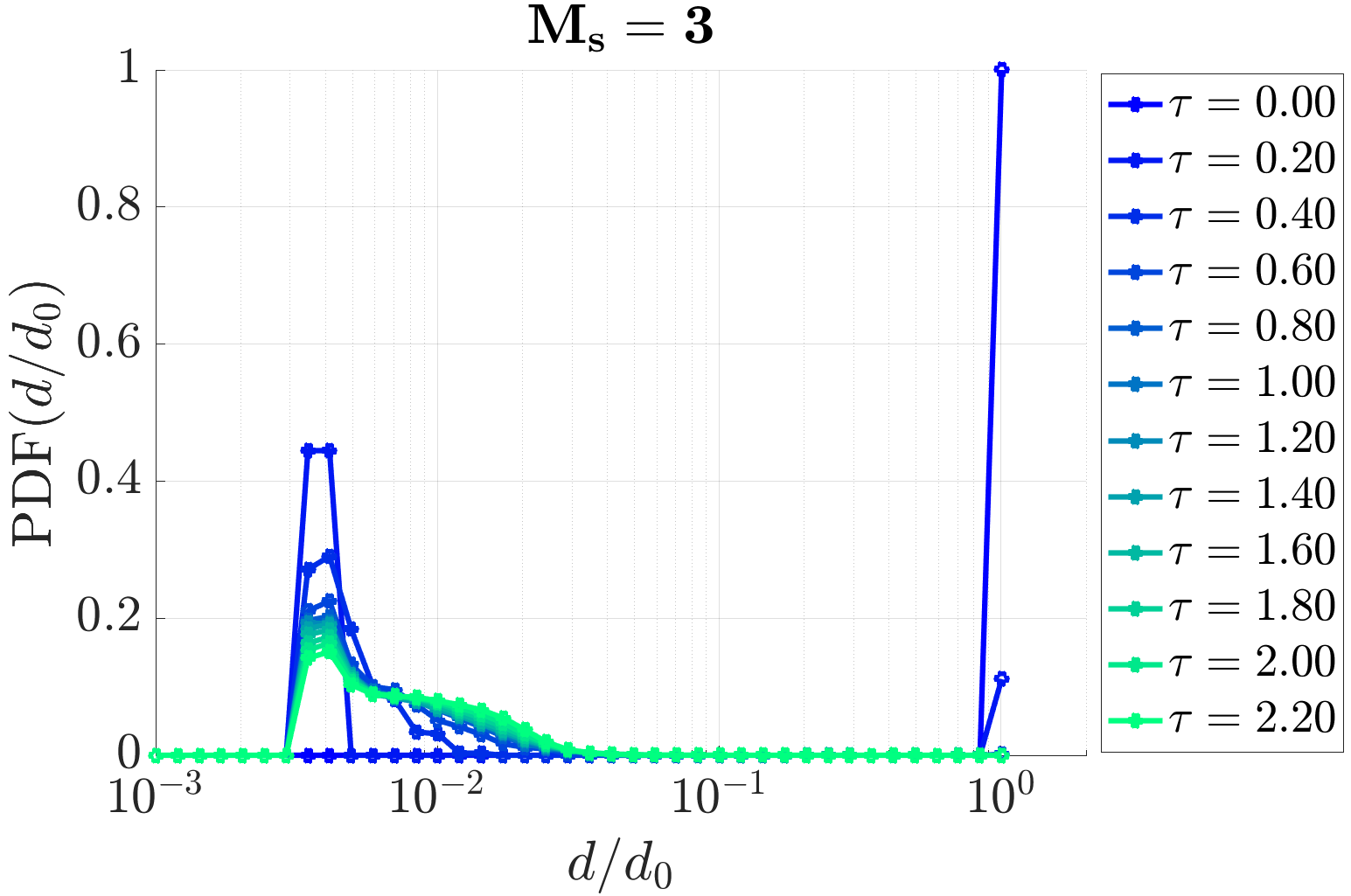} 
     \end{subfigure}
    \caption{PDFs of droplet diameters at selected nondimensional times.}
    \label{fig:d_pdf_tau}
\end{figure}

Constant-$\tau$ cross-sections of the PDFs in Fig.\ \ref{fig:d_pdf_surf_const_taus} are plotted for selected times in Fig.\ \ref{fig:d_pdf_tau}. Here, it is seen that the PDFs of droplet diameter approach bimodal log-normal distributions over time. One of these peaks, centered at $d/d_0 \sim 0.004$, corresponds to the smallest secondary droplets, while the other peak, centered at $d/d_0 \sim 0.01$, corresponds to the medium-sized secondary droplets that become more prevalent as the primary droplet is pierced. It is hypothesized that with further simulation time, these medium-sized droplets would atomize further, yielding a unimodal log-normal distribution for droplet diameter as has been observed in previous work \cite{bielawski_thesis, bielawski2024analysis}. The current simulations were terminated prior to this point due to the growing computational cost associated with capturing the cloud of secondary droplets.


\subsection{Droplet breakup times}

\begin{figure}[h]
    \begin{center}
    \includegraphics[width=0.85\columnwidth]{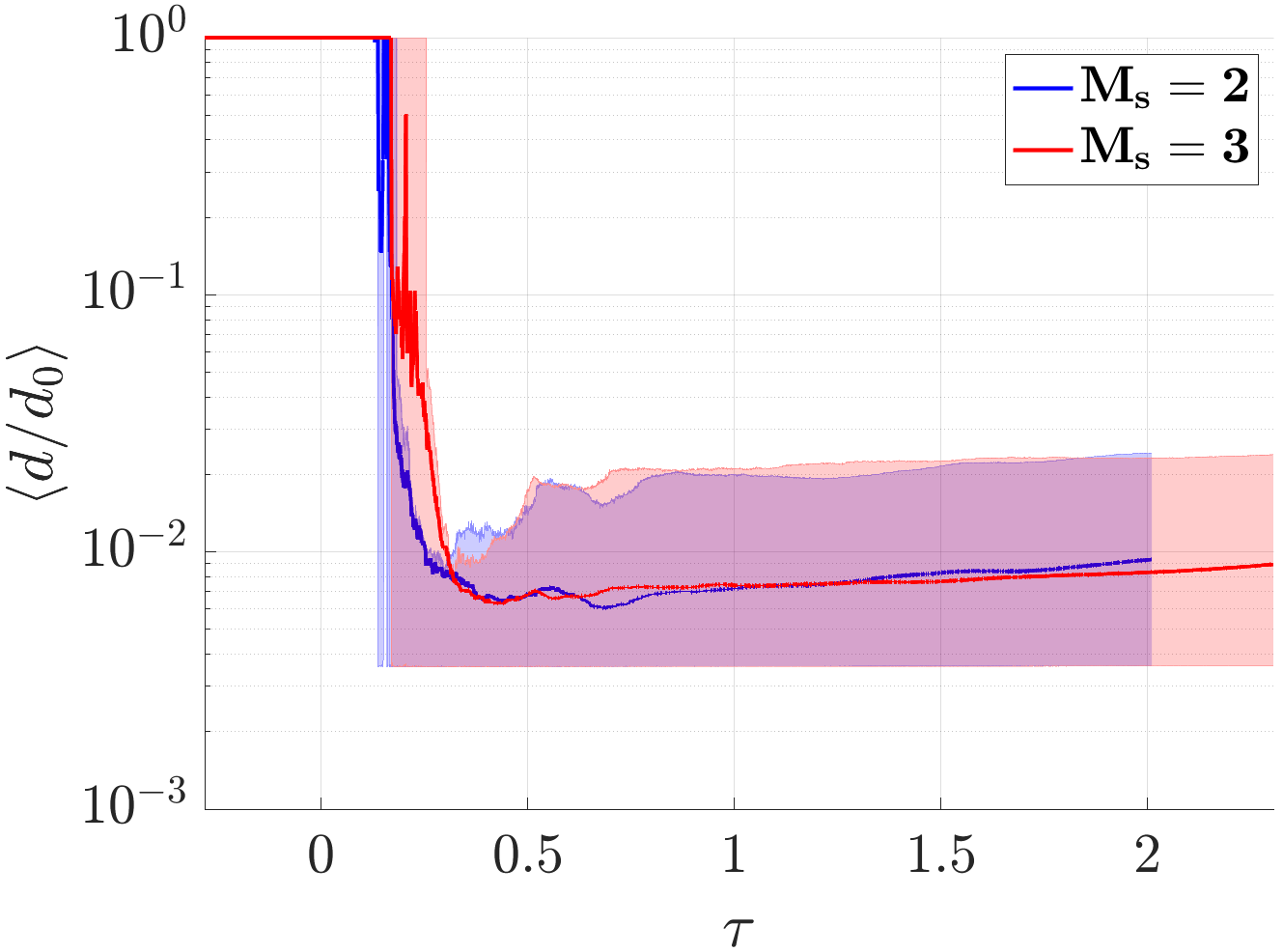}
    \caption{Mean droplet diameters through time for both cases. Error bars show the bounds for 95\% of droplets at a given time.}
    \label{fig:d_avg_tau__combine} 
    \end{center}
\end{figure}

The mean normalized droplet diameters over time are plotted in Fig.\ \ref{fig:d_avg_tau__combine}, with colored bands indicating the range of 95\% of droplet diameters sampled at any given time. Both cases show notable similarities, including the persistence of the primary droplet up to $\tau \sim 0.15$ and means between $0.006 < d/d_0 < 0.01$ for $\tau > 0.4$. After $\tau \sim 0.8$, the mean droplet diameter steadily increases—likely due to increasing numbers of larger secondary droplets that are shed from the pierced primary droplet. This counteracts the prevalence of the smallest secondary droplets, as seen in Figs.\ \ref{fig:d_pdf_surf_const_taus}-\ref{fig:d_pdf_tau}. Notably, in the time span $0.15 < \tau < 0.4$, the mean droplet diameter drops more quickly in the $M_s=2$ case. This indicates faster formation of small secondary droplets, which may be due to the more pronounced ligament formation in the droplet wake (see $\tau=0.386$ in Figs.\ \ref{fig:M2_3D_T_sequence}-\ref{fig:M3_3D_T_sequence}). However, after $\tau \sim 1.3$, the mean droplet diameter is consistently lower in the $M_s=3$ case.

\begin{figure}[!h]
     \centering
     \begin{subfigure}[b]{\columnwidth}
         \centering
         \includegraphics[width=\columnwidth]{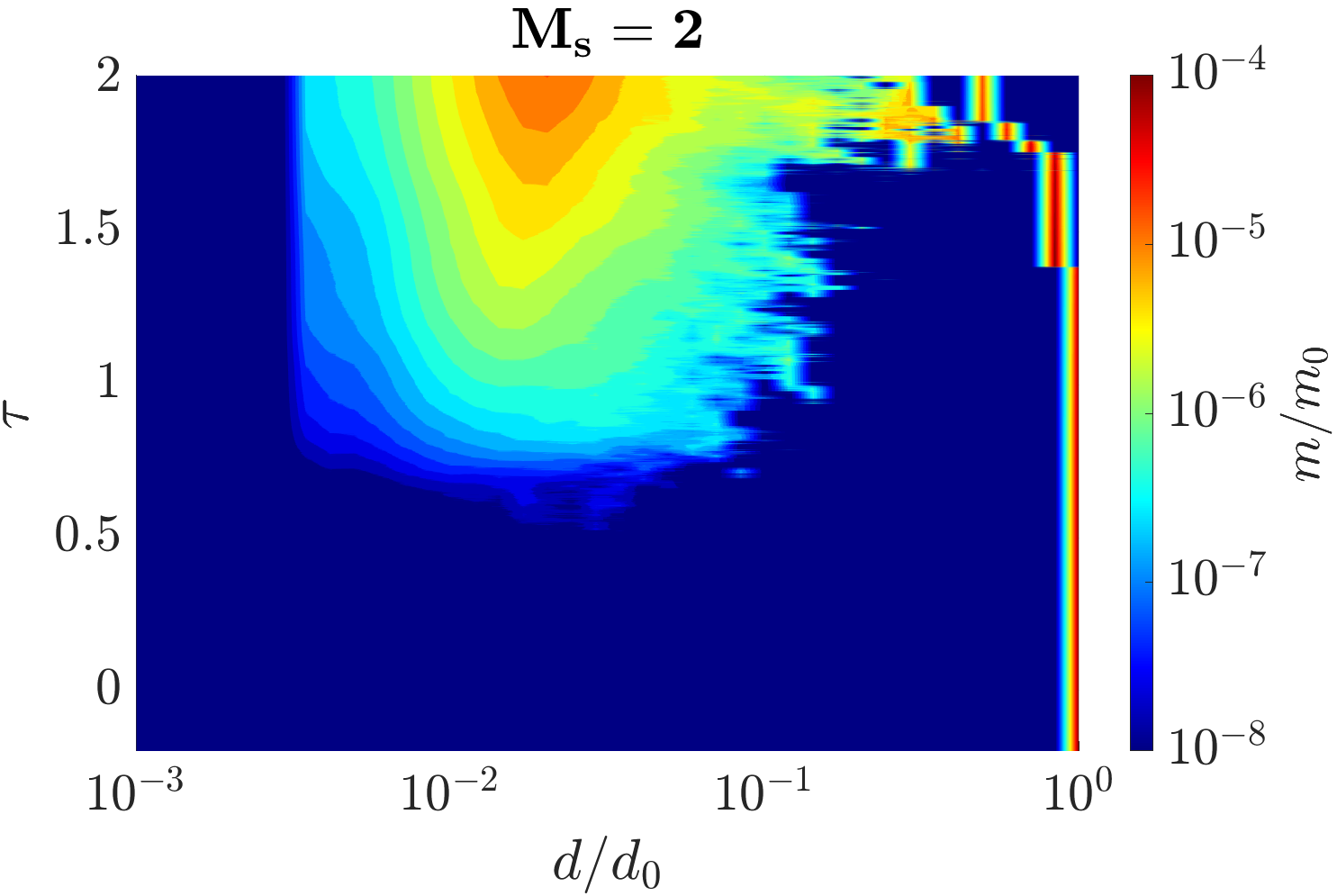} 
     \end{subfigure}
     \\
     \begin{subfigure}[b]{\columnwidth}
         \centering
        \includegraphics[width=\columnwidth]{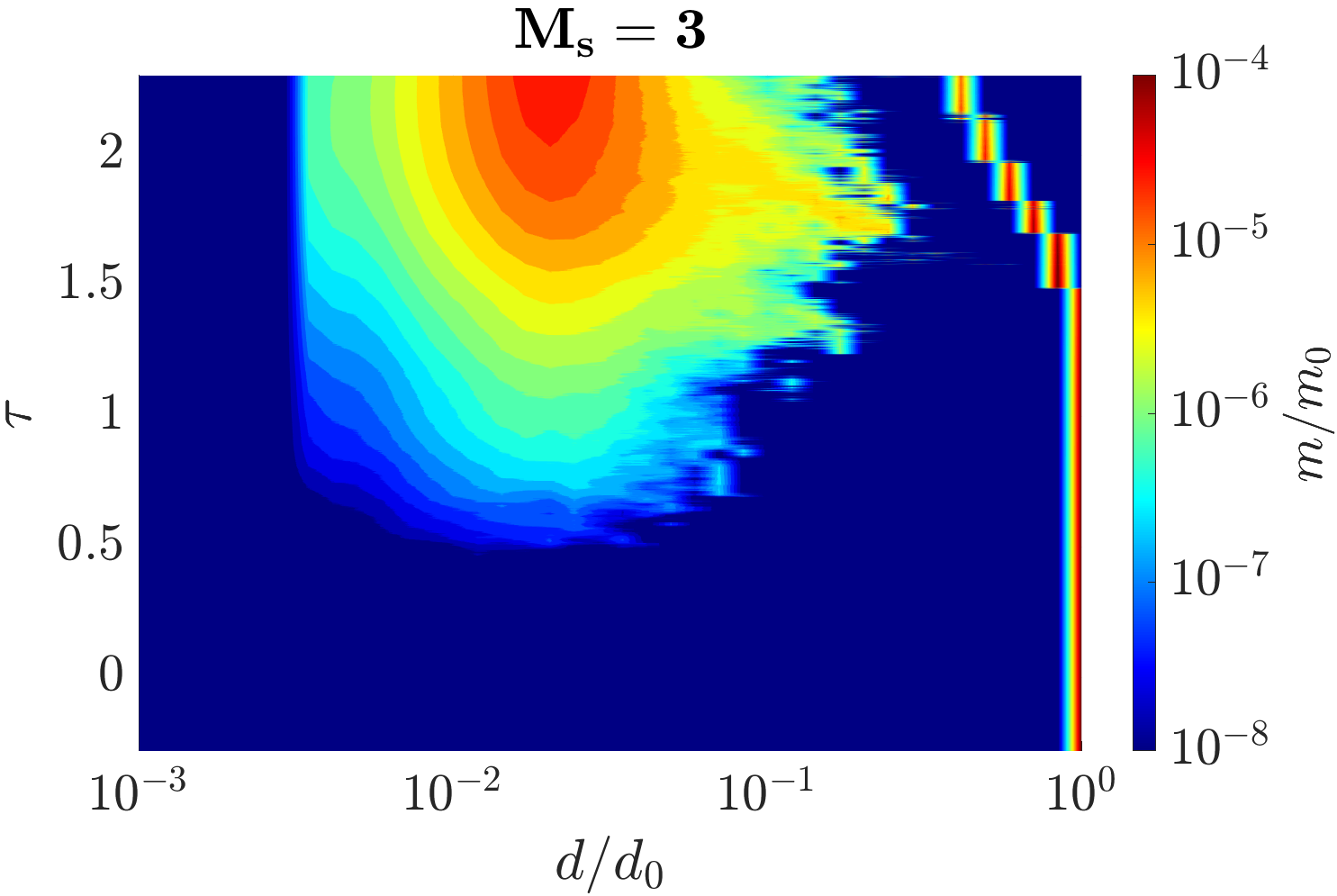} 
     \end{subfigure}
    \caption{Distribution of total droplet mass in various diameter ranges versus nondimensional time. \vspace{-3mm}}
    \label{fig:V_sum_surf}
\end{figure}

The reason for this can be seen in Fig.\ \ref{fig:V_sum_surf}, which shows the total liquid mass in droplets of varying sizes through time. As with the PDFs in Fig.\ \ref{fig:d_pdf_surf_const_taus}, the droplet diameters are grouped into 40 logarithmically-spaced bins, and the total mass of all droplets within each bin are computed at each time. Despite the rapid initial formation of small secondary droplets, which causes the sharp drops in mean droplet diameter in Fig.\ \ref{fig:d_avg_tau__combine}, nearly all of the liquid mass is contained within the primary droplet ($d/d_0 = 1$) prior to $\tau \sim 0.7$. This time of limited mass transfer to smaller droplets has been referred to as the induction time for the droplet in previous work \cite{bielawski_thesis, bielawski2024analysis}. Once the piercing events illustrated in Figs.\ \ref{fig:M2_3D_T_sequence}-\ref{fig:M3_2D_T_sequence} cause the primary droplet to break apart, accelerated atomization leads to more substantial transfer of liquid mass into smaller droplets on the order of 1-10\% of the initial droplet diameter. This atomization occurs more rapidly in the $M_s=3$ case, as can be seen by the higher peaks in the contours for $\tau > 1.5$.

\begin{figure}[h]
    \begin{center}
    \includegraphics[width=0.85\columnwidth]{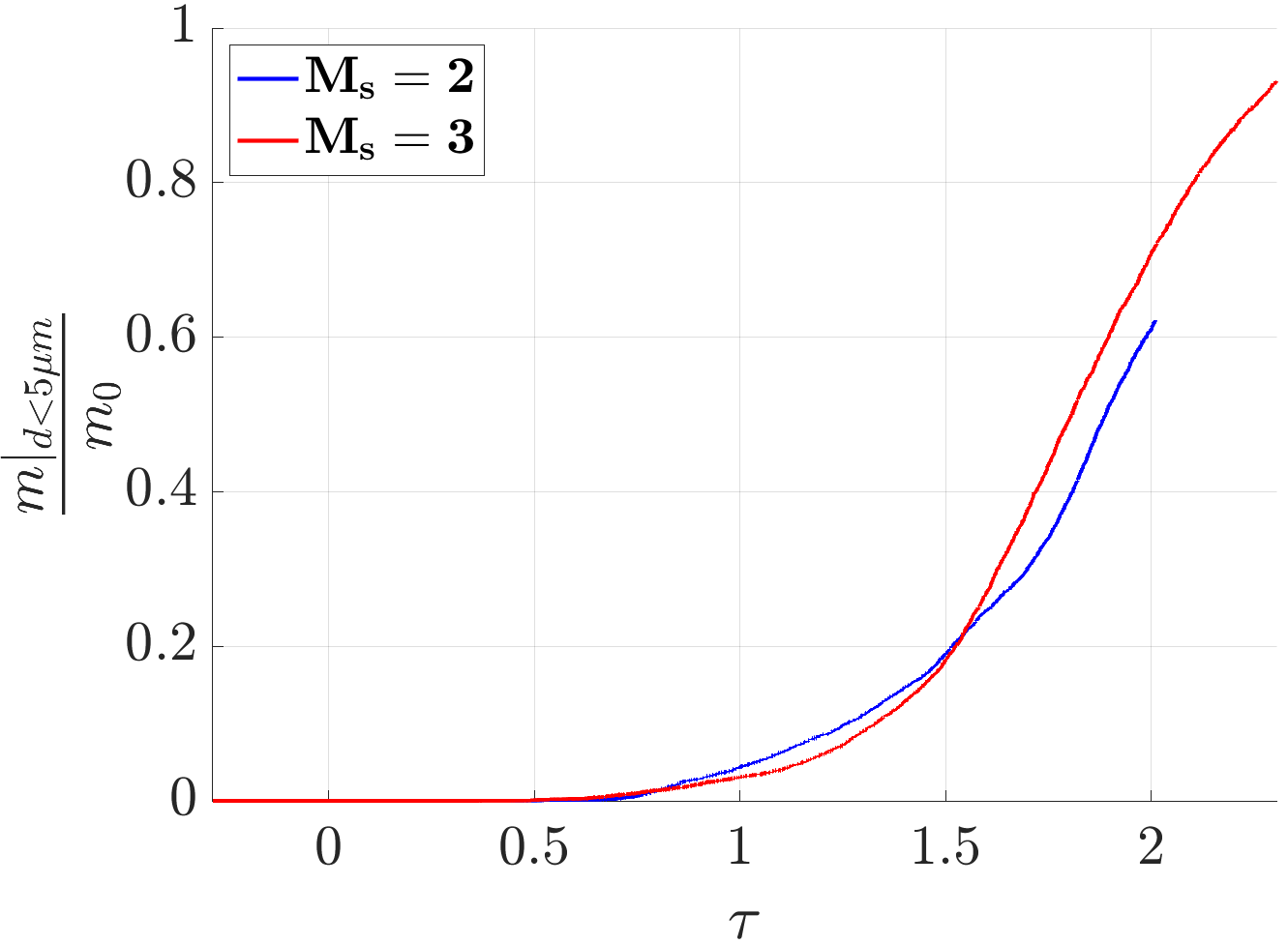}
    \caption{Total mass in droplets with diameter less than 5 $\mu$m normalized by initial droplet mass.}
    \label{fig:V_tot_5um_tau__combine} 
    \end{center}
\end{figure}

This transfer of mass into smaller droplets is critical for liquid-fueled detonations, as their stable propagation requires a sufficient quantity of small fuel droplets that can evaporate and react with the surrounding oxidizer prior to the wave-relative sonic plane. The necessary droplet diameters for this to occur have been estimated as 5-7.5 $\mu$m \cite{schwer2018liquid}. Following this metric, Fig.\ \ref{fig:V_tot_5um_tau__combine} shows the fraction of total liquid mass within droplets less than 5 $\mu$m in diameter over time. This fraction can be interpreted as the degree of atomization of the primary droplet.

As discussed previously, both cases exhibit an induction time of $\tau \sim 0.7$, during which nearly all of the liquid mass is contained within the primary droplet. The degree of atomization is slightly greater between $0.7 < \tau < 1.5$ in the $M_s=2$ case, but following the droplet piercing events at $\tau \sim 1.3$, the rate of atomization in the $M_s=3$ case increases more substantially. This produces a greater degree of atomization at $\tau = 2$ in the $M_s=3$ case (71\%) than in the $M_s=2$ case (61\%). By $\tau = 2.25$, 90\% atomization is achieved in the $M_s=3$ case. The similarities in the slopes of the two profiles for $\tau>1.7$ indicate similar atomization rates during this time frame. This suggests that the discrepancies in the degree of atomization are attributable to differences in the droplet breakup between $0.7 < \tau < 1.7$. As shown in Figs.\ \ref{fig:M2_2D_T_sequence}-\ref{fig:M3_2D_T_sequence}, this time period corresponds to the piercing of the windward droplet surface and the fracturing of its ligaments. Thus, the differences in piercing mode and subsequent droplet morphology may play a role in the increased atomization observed in the $M_s=3$ case.

These breakup times are shorter than those typically observed in experiments of catastrophic droplet breakup. Measurements of nondimensional breakup times include $4 < \tau < 6$ from Ranger and Nicholls \cite{nicholls1969_aiaa}, $\tau=3.5$ from Reinecke and Walden \cite{reinecke1975shock}, $\tau=5.5$ from Hébert et al.\ \cite{hebert2020_shockdroplet}, and $2 < \tau < 5$ from Schroeder et al.\ \cite{schroeder2024_pci}. However, it is worth noting that the discrepancies observed here may be partially attributable to the definition of droplet breakup. In experiments, the breakup time is often defined as the instant at which the droplet or the resultant atomized spray can no longer be distinguished from the background gas. While the results indicate that the primary droplets have largely broken up by $\tau \sim 2$, Figs.\ \ref{fig:M2_3D_T_sequence} and \ref{fig:M3_3D_T_sequence} show that the cloud of secondary droplets can still appear as a cohesive structure when viewed from the side. As such, the shorter breakup times reported here may be due to the definition presented in Fig.\ \ref{fig:V_tot_5um_tau__combine}, which is inaccessible with current experimental diagnostics. The discrepancies could also be partly due to the assumption of quasi-instantaneous thermomechanical equilibrium discussed in the introduction. Finite rate equilibration that takes place over several simulation time steps could lead to a slower deposition of momentum and energy into the droplet, thereby slowing its breakup processes. Further investigation with different numerical approaches would be required to assess these hypotheses.


\section{Conclusions}

This work presented three-dimensional simulations of Mach 2 ($We=822$) and 3 ($We=3760$) shock waves interacting with 100 $\mu$m diameter water droplets. Adaptive mesh refinement was used in a volume of fluid formulation to highly resolve the droplet surface instabilities and secondary atomization. The results show that the breakup processes proceed at similar rates when considering the nondimensional timescale of Range and Nicholls \cite{nicholls1969_aiaa}. Over $\tau<0.5$, the droplet begins to flatten due to the pressure differential across it, and instabilities on the droplet surface form and grow. These instabilities form ligaments and secondary droplets that are shed from the primary droplet, along with facilitating the piercing of the windward droplet surface at $\tau>1$. Differences in the modes of these instabilities lead to different droplet morphologies—a central spike, as in the $M_s=2$ case, or a central piercing, as in the $M_s=3$ case. However, in both cases, the droplet displacement and radial spreading occur at similar rates with respect to $\tau$. Both quantities align well with experimentally-derived empirical relations.

By $\tau = 2$, both cases yield bimodal log-normal PDFs for the secondary droplets' Sauter mean diameters, with peaks centered at $d/d_0 \sim 0.004$ and $d/d_0 \sim 0.01$. It is hypothesized that with longer simulation times, the PDFs would evolve into unimodal log-normal distributions as the secondary atomization proceeded. While small secondary droplets quickly dominate the droplet number density, substantial transfer of liquid mass into smaller droplets is delayed until after the piercing of the primary droplet. The transfer of mass into droplets less than 5 $\mu$m in diameter is initially faster in the $M_s=2$ case, but is surpassed by the $M_s=3$ case for $\tau>1.5$. This leads to 61\% and 71\% atomization by $\tau=2$ in the $M_s=2$ and $M_s=3$ cases, respectively, and roughly 90\% atomization in the $M_s=3$ case by $\tau=2.25$. Similar rates of mass transfer to small droplets are observed for $\tau > 1.7$, suggesting that differences in droplet piercing mode between $0.7 < \tau < 1.7$ may play a role in the increased atomization for $M_s=3$. Additional simulations with different shock strengths and droplet sizes would be required to further ascertain trends in droplet breakup speeds. These trends could then be used to tune empirical models for droplet breakup, which in turn be used in comparatively inexpensive Euler-Lagrangian simulations of more complex systems. Together, these simulations of canonical configurations and full-scale systems can play a valuable role in the development of novel propulsion and power generation technologies based on liquid-fueled detonations.


\begin{acknowledgments}
Support for this research was provided by ONR MURI N00014-22-1-2606, with Dr.\ Steven Martens as program manager. Computational resources were provided by the United States Department of Defense High-Performance Computing Modernization Program (DoD HPCMP). The authors would like to acknowledge the efforts of Shivank Sharma and Lorenzo Angelilli in developing the multiphase solver.

\end{acknowledgments}


\bibliography{main} 

\end{document}